\documentstyle[12pt]{article}
\def\prl{Phys.\ Rev.\ Lett.\ }

\def\prb{Phys.\ Rev.\ B }
\begin{document}
\setlength{\baselineskip}{.375in}
\vspace{1in}
\begin{center}
\Large\bf
Rezayi and Read reply\\
\vspace{0.75in}
\large\sc E. Rezayi${}^1$ and N. Read${}^2$\\
\vspace{0.325in}
\normalsize\em
${}^1$Department of Physics, California State University,\\
Los Angeles, CA 90032\\
and\\
${}^2$Departments of Physics and Applied Physics, P.O. Box 208284\\
Yale University, New Haven, CT 06520\\
\vspace{0.75in}
\end{center}
\begin{verbatim}
PACS Nos. 73.40.Hm, 03.70.+k
March, 1994
\end{verbatim}
\newpage

Jain's Comment \cite{comm} raises two main points of interpretation about our
results \cite{RR}.

1) Jain's first claim is that gapless (or power-law correlated) systems cannot
be addressed by finite size
studies. This is not true. While we agree that finite size systems should be
larger than
{\em some} correlation length, this does not
necessarily rule out gapless or even critical systems. A correlation length
can be defined \cite{amit}
as the scale at which correlation functions approach their asymptotic behavior
{\em even if it is a power law}. At finite sizes larger than this scale, a
power-law correlated quantum system will exhibit discrete energy levels whose
spacing scales as a power of size, and whose structure (scaled
energies versus quantum numbers) exhibits universal properties of the
renormalization-group fixed point governing the infinite system \cite{wilson}.
A well-studied example is the 1-dimensional antiferromagnetic Heisenberg
spin-$\frac{1}{2}$ chain.
(In the usual examples, the low-energy
excitations are at zero wavevector, while in a Fermi liquid they are at $k_F$,
which makes the behavior as a function of size a little more complex,
 but the principle is the same.)

We have studied \cite{RR} the sequence at $N_\phi=2(N-1)$ flux for
$N$ electrons on a sphere for {\em all}
$N<15$, and shown that the structure of the low-lying levels, and the
corresponding wavefunctions, {\em are} what one would
expect in the finite size version of the HLR theory \cite{HLR}.
Because the angular momenta of the
ground states vary with $N$, we have not so far attempted any scaling
of gap sizes with $1/N^{1/2}$ as $N\rightarrow\infty$, though it is clear that
they decrease with increasing size.
Jain, at the beginning and end of his Comment \cite{comm}, seems to assert
that an infinite system is required to study the
compressible $\nu=1/2$ state. We believe that in fact the relevant
correlation length is of order one interelectron spacing, as for other
simple fractions such as $\nu=1/3$.
Jain does not state how he would expect the finite size spectra of a Fermi
liquid of composite fermions to differ from what we see.

2) In the spherical geometry, sequences of finite sizes $(N,N_\phi)$ that
represent finite size approximations to a particular type of ground (or
excited) state at
some $\nu$ always have the form $N_\phi=\nu^{-1}N + C$ for some constant $C$,
and so states that are nominally of different $\nu$ may coincide at the same
$(N,N_\phi)$ if they have different $C$'s.
For two ground states that
occur at different $\nu$ for the same Hamiltonian, there is no
inconsistency {\em in principle} in saying that {\em both} interpretations are
correct when they occur at the same $(N,N_\phi)$. The single spectrum of
the Hamiltonian of the system will evolve in different ways as one passes
to larger systems along the two distinct sequences of sizes.
This occurs in the present studies at
$N=9$, $N_\phi=16$ which simultaneously lies on (at least) three sequences:
$\nu=1/2$ as specified above \cite{RR},
$3/7$ and $3/5$ \cite{comm}. Jain's view is that because he favors the
interpretation as part of the latter two (!) \cite{comm}, it
cannot be part of the former. This is not valid for the reason just given.
Clearly, if one wishes to see the difference between the states, one must go
to larger sizes, as we did at 1/2 \cite{RR}.
One cannot just assert that the system is incompressible by looking at
the spectrum at one size.
Similarly all ``filled shell'' cases $N=n^2$ coincide
with one finite size ground state at each $\nu=n/(2n\pm 1)$. This is not
inconsistent with our interpretation or with the expected reduction of gap
with increasing $N$, which is expected to go as $1/n$ in these
$\nu=n/(2n\pm 1)$ states \cite{HLR} and as $1/N^{1/2}=1/n$ in the Fermi liquid.

However, in terms of composite fermions, {\em Jain's} physical picture
seems implausible when one recalls that at these sizes, as for all
$N_\phi=2(N-1)$, the fermions see {\em zero} net
magnetic field. The wavefunctions are the spherical version of plane
waves \cite{RR}, {\em not Landau level functions as Jain states\cite{comm}}.
Hence, interpreting the $N=9$ state as part of a 3/5 sequence seems
unnatural and finite
size effects may be significant quantitatively at 3/5.
Jain \cite{comm} proposes that $\nu=n/(2n\pm 1)$ states
should be studied at $N\geq n^2$ only; we suspect the true criterion is
$N\gg n^2$. This is supported by ({\it i\/}) the fact that our Fig.\ 3
\cite{RR} does not closely resemble {\em any} spectrum
in \cite{hald} and ({\it ii\/}) the failure of $g(r)-1$ to decay exponentially
\cite{RR} as it would in an incompressible state, so the correlation
length {\em in the 3/5 state} must be larger than this size.
A separate argument, valid also in the periodic boundary condition geometry
(PBCG) where $C$ is always zero, is that the natural correlation length at
$\nu=n/(2n\pm 1)$ is the cyclotron radius for a fermion at the
Fermi surface \cite{HLR},
which diverges as $n$ as $\nu\rightarrow 1/2$. Systems of area $\gg n^2$
are then needed to see {\it e.g.} $N$-independent gaps, while smaller sizes
{\em even at $\nu\neq 1/2$ for PBCG} will more closely resemble the Fermi
liquid with gaps $\sim N^{-1/2}$. Then for all filling factors $\nu=p/q$
considered here, the asymptotic behavior sets in for $N\gg p^2$, which
is well satisfied in our study at $\nu=1/2$ though not in Jain and coworkers'
$N=9$, $\nu=3/5$ study \cite{comm}.

This work was supported by NSF-DMR-9113876 (E.R.) and -9157484 (N.R.).

\vspace*{\fill}
\end{document}